\pdfoutput=1 
\documentclass[12pt]{article}
\usepackage{authblk}
\usepackage{here}
\usepackage{amsmath}
\usepackage{amsfonts}
\usepackage{bm}
\usepackage{graphicx}          
\usepackage{ascmac}
\usepackage{xcolor}
\usepackage{braket}
\usepackage{here}

\usepackage{cite}
\usepackage[top=25truemm,bottom=30truemm,left=20truemm,right=20truemm]{geometry}
\usepackage{afterpage}
\usepackage{cases}
\usepackage{fancybox}
\usepackage{youngtab}

\usepackage[breaklinks=true,pagebackref=true,linktocpage=true]{hyperref}
\hypersetup{colorlinks=true, citecolor=blue, linkcolor=blue, linkbordercolor=0 0 1}
\makeatletter

\@addtoreset{equation}{section}
\makeatother
\newlength{\myh}
\baselineskip=\normalbaselineskip

\newcommand{\Ab}{\bar{A}}
\newcommand{\cAb}{\bar{\cA}}
\newcommand{\Fb}{\bar{F}}

\newcommand{\cA}{{\mathcal A}}

\newcommand{\cF}{{\mathcal F}}

\newcommand{\cL}{{\mathcal L}}

\newcommand{\cO}{{\mathcal O}}

\newcommand{\cR}{{\mathcal R}}

\newcommand{\cY}{{\mathcal Y}}

\newcommand{\bR}{{\mathbb R}}

\newcommand{\bZ}{{\mathbb Z}}

\newcommand{\wt}{\widetilde}

\newcommand{\ha}{\hat{a}}
\newcommand{\hb}{\hat{b}}
\newcommand{\hg}{\hat{g}}
\newcommand{\hA}{\hat{A}}
\newcommand{\hF}{\hat{F}}
\newcommand{\hU}{\hat{U}}

\newcommand{\nn}{\nonumber}

\def\matt[#1,#2,#3,#4]{\left(%
\begin{array}{cc} #1 & #2 \\ #3 & #4 \end{array} \right)}

\DeclareMathOperator{\link}{Link}
\DeclareMathOperator{\Tr}{Tr}
\DeclareMathOperator{\tr}{tr}

\let\ker=\relax
\DeclareMathOperator{\ker}{Ker}
\newcommand{\trr}{\triangleright}

\newcommand{\bb}{\mathbb}
\newcommand{\fr}{\frac}
\newcommand{\ul}{\underline}
\newcommand{\wed}{\wedge}

\newcommand{\er}{\eqref} 
\newcommand{\der}{\partial} 
\renewcommand{\(}{\left(} 
\renewcommand{\)}{\right)}

\renewcommand*{\backref}[1]{}
\renewcommand*{\backrefalt}[4]{%
  \ifcase #1 %
{\bf \color{red}No citations.}
  \or
    (page #2).%
  \else
    (pages #2).%
  \fi%
}

\begin{document}
\title{{\LARGE 
BCF anomaly and higher-group structure in
the low energy effective theories of  mesons
}}
\author[1]{Tatsuki Nakajima}
\author[1,2]{Tadakatsu Sakai}
\author[3]{Ryo Yokokura}
\affil[1]{Department of Physics, Nagoya University}
\affil[2]{Kobayashi-Maskawa Institute for the Origin of Particles 
and the Universe, Nagoya University}
\affil[3]{
KEK Theory Center, Tsukuba 305-0801, Japan,
\par 
Research and Education Center for Natural Sciences,
Keio University, Hiyoshi 4-1-1, Yokohama, Kanagawa 223-8521, Japan}

\date{}
\maketitle
\thispagestyle{empty}
\setcounter{page}{0}

\begin{abstract}
We discuss the BCF anomaly of massless QCD-like theories, first 
obtained by Anber and Poppitz, from the viewpoint of the
low energy effective theories. We assume that the QCD-like theories
exhibit spontaneous chiral symmetry breaking due to a quark 
bilinear condensate. 
Using the 't Hooft anomaly matching condition for the BCF anomaly,
we find that the low energy effective action is composed
of a chiral Lagrangian and a Wess-Zumino-Witten term together
with an interaction term of the $\eta^\prime$ meson with
the background gauge field for a discrete one-form symmetry.
It is shown that the low energy effective action cancels
the quantum inconsistencies associated with $\eta^\prime$ 
due to an ambiguity of how to
uplift the action to a five-dimensional spacetime with a boundary.
The $\eta^\prime$ term plays a substantial role in 
exploring the emergent higher-group structure
at low energies.

\end{abstract}

\setcounter{section}{+0}
\setcounter{subsection}{+0}

\newpage

\afterpage{\clearpage}
\newpage
%
%
\tableofcontents
\section{Introduction}

It has been established that 
higher-form symmetries~\cite{gensym} (see also Refs.~\cite{Batista:2004sc,Pantev:2005zs,Pantev:2005wj,Nussinov:2006iva,Nussinov:2008aa,Nussinov:2009zz,Nussinov:2011mz,Banks:2010zn,Distler:2010zg,Kapustin:2013uxa,kapsei})
provides 
us with a powerful tool for studying nonperturbative aspects of
quantum field theory~\cite{Gaiotto:2017yup,Gaiotto:2017tne,Tanizaki:2017qhf,Tanizaki:2017mtm,Komargodski:2017dmc,Hirono:2018fjr,Hirono:2019oup,Hidaka:2019jtv,Misumi:2019dwq,Tanizaki:2022ngt}.
A first attempt to utilize higher-form symmetries to quantum
chromodynamics(QCD)
with massless flavor degrees of freedom is made in \cite{shiyon}
on the basis of a careful analysis of faithful group action of 
the symmetries for the case of a certain ratio of the number of 
the flavors to that of the colors.
The paper \cite{anbpop1} studies QCD-like theories of
gauge group $SU(N_c)$ with the quark fields 
belonging to an irreducible representation $R$.
It is found there that the QCD-like theories admit discrete one-form
symmetries for generic $R$ and flavors as well. 
{}Furthermore, turning on the background gauge fields for the
flavor and one-form symmetries, the associated 't Hooft anomaly
is worked our and called the baryon-color-flavor(BCF) anomaly.
{}For recent studies of the BCF anomaly,
see \cite{anbpop2, Anber:2020gig, Anber:2021lzb, Anber:2021iip}.
In particular, by coupling the QCD-like theories with a neutral and
complex Higgs field,
the paper \cite{anbpop2} focuses on an RG flow to an axion system
that is triggered by the Higgs vev.
It is shown that the low energy effective theory is given 
by a BF-type action and reproduces the BCF anomaly as expected from
the 't Hooft anomaly matching condition.
It is also pointed out that three-group structure is manifested as
a Green-Schwarz(GS) transformation 
\cite{Green:1984sg} of a dynamical three-form gauge 
potential under a discrete one-form gauge transformation.
It is known that
the mathematical structure of higher-form symmetries is naturally
formulated in terms of higher-group.
Roughly speaking, the higher-group is a set of the 
groups that describe higher-form symmetries
including correlations among them.
{}For recent developments in understanding the QFT dynamics from 
the viewpoint of higher-group structure, see 
\cite{Tachikawa:2017gyf, Cordova:2018cvg, Benini:2018reh, Delcamp:2018wlb,Delcamp:2019fdp, Hsin:2019fhf,Tanizaki:2019rbk, Hidaka:2020iaz, Cordova:2020tij, Hidaka:2020izy,Hidaka:2021mml,Hidaka:2021kkf,Nakajima:2022feg} for instance.

It is a long-standing problem to determine the phase structures of
the QCD-like theories such as spontaneous chiral symmetry breaking, 
unbroken chiral symmetry leading to massless baryons, conformal window, etc.
In this paper, we assume that the QCD-like theories exhibit
spontaneous chiral symmetry breaking. Then, the low energy effective 
theory is given by nonlinear $\sigma$-models that are constructed
from nonlinear realization of the associated chiral symmetry
breaking.
The 't Hooft anomaly matching condition for the chiral flavor symmetry
requires that the Wess-Zumino-Witten(WZW) term \cite{Wess:1971yu,Witten:1983tw} 
be added to the nonlinear $\sigma$-model.
It is found that 
a naive generalization of the WZW term to the cases
in the presence of the BCF background fields is insufficient 
because the BCF anomaly contains a mixed 't Hooft anomaly between
a discrete axial symmetry and a discrete one-form symmetry.
It is argued that the mixed anomaly
can be reproduced by adding
an interaction term between a $U(1)$ meson and a background three-form
gauge field.
This term is not left unchanged under a one-form gauge symmetry 
transformation. It is found that the one-form symmetry
is restored by introducing a background three-form gauge field
that makes a GS-type transformation under the
one-form gauge symmetry transformation.
This is a manifestation of three-group structure in the low energy
effective action of the QCD-like theories.

We also examine a quantum inconsistency associated with the $U(1)$ meson, 
which is realized as an ambiguity
of how the low energy effective action is uplifted to a five-dimensional 
one that has manifest gauge invariance.
We call it an operator-valued ambiguity. This is required to vanish
for the consistency of the quantum QCD-like theories.
It is discussed that in the presence of the BCF background gauge fields,
both the WZW term and the $U(1)$ meson 
term suffer an operator-valued ambiguity. However, no ambiguity arises in 
the total effective action because of cancellation of the
ambiguities from the two terms.

The organization of this paper is as follows.
In section 2, we review the BCF anomalies in the QCD-like
theories.
Section 3 is devoted to a derivation of the low energy effective action
of the QCD theories that reproduces the BCF anomaly, assuming that
a quark bilinear condensate gives rise to spontaneous chiral symmetry 
breaking. We also show cancellation of the operator-valued ambiguity associated
with the $U(1)$ meson. We end this paper with discussions in
section 4.
In appendix \ref{leet:Rr}, we demonstrate in detail
how to obtain the low energy effective action when the quark fields
belong to a real representation of $SU(N_c)$.

\section{Review of BCF anomaly}

In this section, we review the BCF anomaly \cite{anbpop1}.

We consider a QCD-like $SU(N_c)$ gauge theory coupled with 
massless fermions
in $(3+1)$-dimensional spacetime.
The matter content of the QCD-like theory we analyze reads
\begin{align}
  \begin{array}{cccc}
&SU(N_c) &U(N_f)_L    &U(N_f)_R \\
\psi_{L\alpha}^{\,ai}& R&  \Box_1
 & 1_0
\\
\psi_{R}^{\dot{\alpha}ai^{\prime}}& R & 1_0 &  \Box_{1}
\\
a& {\rm\bf adj.} &1_0&1_0
  \end{array}
\label{modelR}
\end{align}
Here, $(\psi_{L\alpha}^{\,ai},\psi_{R}^{\dot{\alpha}ai^{\prime}})$
are fermions, and 
$a = a_\mu dx^\mu$ is 
an $\frak{su}(N_c)$-valued 
1-form gauge field.
$R$ 
denotes a representation of $SU(N_c)$ to which 
the fermions belong.
${\bf adj.}$ means the adjoint representation of $SU(N_c)$.
The undotted and dotted 
indices $\alpha,\dot{\alpha}$ are
 the left- and right-handed 
spinor indices
for the Lorentz symmetry,
$\alpha, \dot\alpha = 1,2$.
The index $a$ runs over the dimension of
the gauge group $SU(N_c)$,
$a=1,2,\cdots, \dim R$.
The indices $i, i^\prime = 1,..., N_f$ are the chiral flavor indices of
$U(N_f)_L$ and $U(N_f)_R$, respectively.
The action is given by
\begin{equation}
 S = \int  d^4 x \(
-\fr{1}{2g^2} \tr f_{\mu\nu} f^{\mu\nu } 
+  \bar\psi  (i\der_\mu + a^R_\mu ) \psi 
+ \fr{\theta }{32\pi^2} \epsilon^{\mu\nu\rho\sigma} 
\tr f_{\mu\nu} f_{\rho\sigma}\),
\quad
 \psi = 
\(
\begin{matrix}
\psi_{L \alpha}
\\
\psi_R^{\dot\alpha} 
\end{matrix}
\).
\end{equation}
Here, $a^R$ is the dynamical $SU(N_c)$ gauge field that couples
with the matter fields in the representation $R$.
It may be complex (e.g., fundamental) or real (e.g., adjoint)
representation.
When $R$ is real, the chiral flavor symmetry enhances as
\begin{align}
  U(N_f)_L\times U(N_f)_R \to U(2N_f) \ .
\end{align}
This is because we can define 
an undotted spinor $\Psi$ as 
\begin{align}
  \Psi_{\alpha}^{aI}=(\psi_{L\alpha}^{ai},\,
\bar{\psi}_{R\alpha i^{\prime}}^{\,a}) \ ,~~~
(I=1,2,\cdots,2N_f)
\end{align}
where
\begin{align}
\bar\psi_{Ri^\prime}^{\,\alpha a}
= \left(\psi_R^{\dot{\alpha}ai^{\prime}}\right)^{\ast}
=\epsilon^{\alpha\beta}\,\bar\psi_{R\beta i^\prime}^{\,a} \ .
\end{align}
This fermion $\Psi$ belongs to the defining representation of $U(2N_f)$.
$U(1)_V$ is identified with the Cartan part of $SU(2N_f)$ with the
generator given by $I_{N_f}\oplus(-I_{N_f})$.
$U(1)_A$ is the $U(1)$ subgroup of $U(2N_f)$ that is orthogonal to
the semi-simple $SU(2N_f)$.

\subsection{ABJ and mixed 't Hooft anomalies}

We turn on the background gauge fields for the flavor subgroup 
$SU(N_f)_V\times U(1)_B\times U(1)_A$,
which are denoted by $A$, $A_B$ and $\chi$, respectively.
Here, $SU(N_f)_V$ is the vector-like
subgroup of $SU(N_f)_L\times SU(N_f)_R$, and
$U(1)_L \times U(1)_R =(U(1)_B \times U(1)_A)/\bZ_2$ with
$\bZ_2$ acting as
\begin{align}
  \bZ_2:~(e^{i\alpha_V},e^{i\gamma_5\alpha_A})
\in U(1)_V\times U(1)_A \to
-(e^{i\alpha_V},e^{i\gamma_5\alpha_A}) \ .
\nn
\end{align}
Then, the fermionic part of the Lagrangian becomes
\begin{align}
\cL_f&
=\bar{\psi}\gamma^\mu\left(
i\partial_\mu+a^R_\mu\otimes I_{N_f}
+I_{d(R)}\otimes A_\mu-(A_{B\mu}-\gamma_5\chi_\mu)\,I_{d(R)}\otimes I_{N_f}
\right) \psi 
\nn\\
&=
\bar{\psi}\gamma^\mu\left(
i\partial_\mu+\cL_\mu\right)P_L\, \psi 
+
\bar{\psi}\gamma^\mu\left(
i\partial_\mu+\cR_\mu\right)P_R\, \psi 
\ ,
\end{align}
where $P_{L,R}=(1\mp\gamma_5)/2$ and
\begin{align}
  \cL&=a^R\otimes I_{N_f}
+I_{d(R)}\otimes A-(A_{B}+\chi)\,I_{d(R)}\otimes I_{N_f} 
= a^R\otimes I_{N_f}
+I_{d(R)}\otimes \Ab_L  \ ,
\nn\\
  \cR&=a^R\otimes I_{N_f}
+I_{d(R)}\otimes A-(A_{B}-\chi)\,I_{d(R)}\otimes I_{N_f}
= a^R\otimes I_{N_f}
+I_{d(R)}\otimes \Ab_R  
 \ .
\end{align}
Here,
\begin{align}
  \Ab_L=A-(A_B+\chi)I_{N_f}=\Ab-\chi I_{N_f} \ ,~~~
  \Ab_R=A-(A_B-\chi)I_{N_f}=\Ab+\chi I_{N_f} \ ,
\label{AbLR}
\end{align}
with $\Ab=A-A_BI_{N_f}$ being the $SU(N_f)_V\times U(1)_B$ gauge field.
$I$ is an identity matrix of rank equal to the suffix and
$d(R)={\rm dim}R$.
We regard $a^R$ as a background gauge field for the moment and
define the effective action
\begin{align}
  e^{i\Gamma[\cL,\cR]}
=\int D\psi D\bar{\psi}\,e^{i\int d^4x\,\cL_f} \ .\nn
\end{align}
The consistent chiral anomaly, which is written in the L-R form,
reads
\begin{align}
  \delta\Gamma=\frac{-1}{24\pi^2}
\int \Tr \left[
\lambda_L\,d\left(\cL d\cL-\frac{i}{2}\cL^3\right)
-
\lambda_R\,d\left(\cR d\cR-\frac{i}{2}\cR^3\right)
\right] \ ,
\end{align}
where the infinitesimal gauge transformations for $\cL$ and $\cR$ are
given by
\begin{align}
  \delta \cL=d\lambda_L-i\cL\lambda_L+i\lambda_L\cL \ ,~~
  \delta \cR=d\lambda_R-i\cR\lambda_R+i\lambda_R\cR \ .
\nn
\end{align}
Here, the symbol ``$\Tr$'' is a trace taken over the product space of the 
representation $R$ of ${\mathfrak{su}}(N_c)$ and the
fundamental representation of ${\mathfrak{su}}(N_f)$.
The consistent anomaly can be rewritten into a covariant form by
adding the local counter term \cite{bz,par}
\begin{align}
\cY[\cL,\cR]
&=
\frac{-1}{48\pi^2}
\int\tr \left[
(F_{\cR}+F_{\cL})(\cR\cL-\cL\cR)
+i(\cR^3\cL-\cL^3\cR)
-\frac{i}{2}\cR\cL\cR\cL
\right] \ ,
\end{align}
with
\begin{align}
  F_{\cL}=d\cL-i\cL^2 \ ,~~
  F_{\cR}=d\cR-i\cR^2 \ . \nn
\end{align}
This makes the vector-like flavor symmetry anomaly free.
It is easy to verify that
\begin{align}
  \cY[\cL,\cR]
=\cY_c[a,\chi]+\cY_f[\Ab,\chi] \ .
\end{align}
Here, 
\begin{align}
 \cY_c=-\frac{2N_fT(R)}{24\pi^2}\int \chi
\tr\left(4fa+ia^3\right) \ ,~~
 \cY_f=-\frac{d(R)}{24\pi^2}\int \chi
\tr\left(4\Fb\Ab+i\Ab^3\right) \ .
\label{cYcf}
\end{align}
$a$ is the dynamical $SU(N_c)$ gauge field for $R=\Box$, and
with
\begin{align}
  f=da-ia^2\ ,~~\Fb=d\Ab-i\Ab^2 \ .\nn
\end{align}
Here, the symbol ``$\tr$'' is a trace taken over the fundamental 
representation of either ${\mathfrak{su}}(N_c)$ 
or ${\mathfrak{su}}(N_f)$.
Throughout this paper, we assume $d\chi=0$, which suffices 
for later purposes.
With the local counter term $\cY$, the chiral anomaly 
takes the form
\begin{align}
  \delta(\Gamma+\cY)
&=
\frac{1}{4\pi^2}\int 
\Big[
2T(R)\,\tr(\hat{\alpha})\tr (f\wedge f)
+d(R)\,\tr (\hat{\alpha}\,\Fb\wedge\Fb)
\Big] \ .
\end{align}
Here, $\hat{\alpha}$ is a $U(N_f)_L\times U(N_f)_R$ rotation 
angle given by
\begin{align}
\frac{1}{2}(\lambda_R-\lambda_L)=I_{d(R)}\otimes \hat{\alpha} \ .
\label{halpha}
\end{align}
The $U(1)_A$ rotation angle $\alpha_A$ is equal to the
trace part of $\hat{\alpha}$:
\begin{align}
  \hat{\alpha}=\alpha_AI_{N_f}+\alpha \ ,~~~\tr\alpha=0 \ .
\label{def:alphaA}
\end{align}
The chiral anomaly for the $U(1)_A$ transformation
is given by
\begin{align}
  \delta_A(\Gamma+\cY)
&=
\frac{1}{4\pi^2}\int \alpha_A
\Big[
2N_fT(R)\,\tr (f\wedge f)
+d(R)\,\tr (\Fb\wedge\Fb)
\Big]
\nn\\
&=
\frac{1}{4\pi^2}\int \alpha_A
\Big[
2N_fT(R) \tr(f\wedge f)
+d(R) \tr(F\wedge F)
+N_fd(R)\,F_B\wedge F_B
\Big] \ .
\label{dGdY}
\end{align}

It follows from the quantization condition
\begin{align}
  \frac{1}{8\pi^2}\int\tr (f\wedge f) \in \bZ \ ,
\nn
\end{align}
that 
the Adler-Bell-Jackiw(ABJ) anomaly \cite{Adler:1969gk, Bell:1969ts}
leaves $\bZ_{4N_fT(R)}\subset U(1)_A$ unbroken such that
\begin{align}
  \alpha_A=\frac{2\pi n_A}{4N_fT(R)} \ ,
\end{align}
with $n_A \in\bZ \mbox{ mod } 4N_fT(R)$.
We also note that the subgroup $\bZ_2\subset\bZ_{4N_fT(R)}$ acts as
\begin{align}
(\psi_L,\psi_R)\to -(\psi_L,\psi_R) \ ,
\end{align}
which is an element of the vector-like $U(1)_B$. As found above,
$\cY$ makes the vector-like flavor group anomaly free so that
the anomaly free $U(1)_A$ subgroup is given by
$\bZ_{2N_fT(R)}=\bZ_{4N_fT(R)}/\bZ_2$ with $n_A\in\bZ\mbox{ mod }2N_fT(R)$.

As we will see shortly, the BCF anomaly is obtained by
incorporating the background gauge potentials for
one-form symmetries into (\ref{dGdY}) in a manner
consistent with the gauge symmetries.

\subsection{BCF anomaly}
\label{S:BCF}

We have discussed conventional anomalies between 
axial, baryon, and flavor symmetries.
In addition, this system can have a $\bb{Z}_{N_c}$
 center symmetry for the gauge group 
$SU(N_c)$ if the fermions belong to a real representation of $SU(N_c)$.
The center symmetry for the gauge group can be recently understood
as a one-form symmetry associated to a 
$\bb{Z}_{N_c}$ rotation of a Wilson loop
in the fundamental representation.
Furthermore, we can also have one-form symmetries even if 
the fermions belong to the (anti-)fundamental representation 
using a simultaneous rotation of the baryon symmetry.
The one-form symmetry also have a mixed 't Hooft anomaly with
the baryon and flavor symmetries, which is called the BCF anomaly.

To work out the anomaly, we need to specify faithful group action
of the symmetries on the quark fields $\psi_L$ and $\psi_R$.
As noted in \cite{anbpop1}, 
it is given by
\begin{align}
  \frac{SU(N_c)\times SU(N_f)_V\times U(1)_B}{\bZ_{N_c}\times \bZ_{N_f}} \ ,
\label{bcf_group}
\end{align}
where the identifications of the elements 
in $SU(N_c)\times SU(N_f)_V \times U(1)_B$ by 
the discrete group $\bZ_{N_c}\times \bZ_{N_f}$ 
are given by
\begin{align}
  \bZ_{N_c}:&~(g,g_V,g_B)\to (e^{2\pi i/N_c}g,g_V,e^{-2\pi i n/N_c}g_B) \ ,
\nn\\
  \bZ_{N_f}:&~(g,g_V,g_B)\to (g,e^{2\pi i/N_f}g_V,e^{-2\pi i/N_f}g_B) \ ,
\label{identification}
\end{align}
with
\begin{align}
  (g,g_V,g_B) \in SU(N_c)\times SU(N_f)_V\times U(1)_B \ .
\end{align}
Here, the elements 
$g$ and $g_V$ are realized in the fundamental representation.
The quantity $n$ is the $N$-ality of the representation $R$.
Let $g^R$ be the $SU(N_c)$ transformation matrix in the representation
$R$. Then, 
$g^R\otimes g_V\otimes g_B$ is left invariant under 
$\bZ_{N_c}\times \bZ_{N_f}$, showing that
the quotient (\ref{bcf_group}) leads to
faithful group action on
$\psi_{L}$ and $\psi_R$.

Now, we define the gauge fields for (\ref{bcf_group}).
They are obtained 
by starting with the direct product of the $SU(N_c)/\bZ_{N_c}$ and
$SU(N_f)_V/\bZ_{N_f}$ bundles,
and then requiring the constraints to be explained shortly.
As discussed in \cite{kapsei}, the $SU(N_c)/\bZ_{N_c}$ and
$SU(N_f)_V/\bZ_{N_f}$ bundles are
constructed by 
lifting $a$ and $A$ to a $U(N_c)$ and a $U(N_f)$ gauge fields
\begin{align}
  \hat{a}=a+\frac{1}{N_c}CI_{N_c} \ ,~~~
  \hat{A}=A+\frac{1}{N_f}C_V\,I_{N_f} \ ,
\label{UNcUNf}
\end{align}
respectively.
$\ha$ and $\hA$ are realized in the fundamental representations of
$\frak{u}(N_c)$ and $\frak{u}(N_f)$, respectively.
$C $ and $C_V$ are one-form gauge fields which are 
assumed to be properly normalized by the flux quantization 
conditions on a closed surface ${\cal S}$,
\begin{equation}
 \int_{\cal S} dC\ , ~\int_{\cal S} dC_V \in 2 \pi \bb{Z}\ .
\end{equation}
The field strengths for $\hat{a}$ and $\hat{A}$ 
are defined as
\begin{equation}
 \hat{f} = d \hat{a} -i \hat{a} \wed \hat{a}\ ,
\quad
\hat{F} = d\hat{A} -i \hat{A} \wed \hat{A}\ . 
\end{equation}
Note that $a $, $\fr{1}{N_c} C$, $A$, and $\fr{1}{N_f} C_V$ in \er{UNcUNf}
are not properly normalized one-forms, but 
$\hat{a}$ and $\hat{A}$ are properly normalized,
\begin{equation}
\fr{1}{2\pi} \int_{\cal S} \tr \hat{f} \in  \bb{Z}\ ,
\quad
\fr{1}{8\pi^2} 
\int \tr (\hat{f} \wed \hat{f}) \in \bb{Z}\ ,
\quad 
\fr{1}{2\pi} \int_{\cal S} \tr \hat{F} \in  \bb{Z}\ ,
\quad
\fr{1}{8\pi^2} 
\int \tr (\hat{F} \wed \hat{F}) \in \bb{Z}\ ,
\end{equation}
on a spin manifold.

$SU(N_c)/\bb{Z}_{N_c}$
and $SU(N_f)/ \bb{Z}_{N_f}$ gauge fields are obtained
by considering $U(N_c)$ and $U(N_f)$ gauge fields 
that are defined to 
obey the gauge transformation laws at
double overlaps of the coordinate patches:
\begin{align}
  \hat{a}_j=\hg_{ji}\,\hat{a}_i\,\hg_{ji}^{-1}
+i\hg_{ji}\,d\hg_{ji}^{-1}+\Lambda_{ji}I_{N_c} \ ,
~~~
  \hat{A}_j=\hg_{Vji}\,\hat{A}_i\,\hg_{Vji}^{-1}
+i\hg_{Vji}\,d\hg_{Vji}^{-1}+\Lambda_{Vji}I_{N_f} \ .
\label{ahat_ij:fund}
\end{align}
This implies that $C$ and $C_V$ transform as
\begin{align}
  C\to C+N_c\Lambda \ ,~~~
  C_V\to C_V+N_f\Lambda_V \ .
\label{Ctransf}
\end{align}
The gauge fields $\hat{a}$ and $\hat{A}$ 
can be regarded as $SU(N_c)/ \bb{Z}_{N_c}$ 
and $SU(N_f)_V/\bb{Z}_{N_f}$ gauge fields 
respectively because
(\ref{Ctransf}) gauges away
$U(1)$ parts of the $U(N_c)$ and $U(N_f)$ gauge fields.

The transformation law (\ref{ahat_ij:fund})
can be extended to that of a gauge potential for the representation
$R$ of $\mathfrak{u}(N_c)$:
\begin{align}
  \hat{a}_j^R=\hg^R_{ji}\,\hat{a}_i\,\hg_{ji}^{R-1}
+i\hg^R_{ji}d\hg_{ji}^{R-1}+n\Lambda_{ji}I_{d(R)} \ ,
\end{align}
with $g^R$ being a $U(N_f)$ gauge transformation for
the representation $R$.
{}For the purpose of constructing the gauge fields for \eqref{bcf_group},
we consider 
\begin{align}
  \wt A=\hat{a}^R\otimes I_{N_f}
+I_{d(R)}\otimes\hat{A}-\hat{A}_B\,I_{d(R)}\otimes I_{N_f}
\label{BCFgauge}
\end{align}
We require that 
$\wt{A} $ obey the gauge transformation law,
\begin{align}
  \wt A_j=(\hg^R\hg_Vg_B)_{ji}\,\wt A_i\,(\hg^R\hg_Vg_B)_{ji}^{-1}
+i(\hg^R\hg_Vg_B)_{ji}\,d(\hg^R\hg_Vg_B)_{ji}^{-1} \ ,
\label{BCFtransf}
\end{align}
or equivalently,
\begin{align}
  \wt A \to 
(\hg^R \hg_V g_B)\wt A\,(\hg^R\hg_Vg_B)^{-1}
+i\hg^R d \hg^{R-1}
+i\hg_V  d \hg_V^{-1}
+ig_B  d g_B^{-1} \ .
\end{align}
This is achieved by setting
\begin{align}
  \hat{A}_B: =A_B+\frac{n}{N_c}C+\frac{1}{N_f}C_V \ ,
\end{align}
which shows that $\hA_B$ transforms
under \er{Ctransf} as
\begin{align}
  \hat{A}_B\to\hat{A}_B+n\Lambda+\Lambda_V\ .
\label{220124.1156}
\end{align}
$\wt A$ couples to the fermion $\psi$ in a gauge invariant manner, and
hence the group (\ref{bcf_group}) is a symmetry 
of the theory (\ref{modelR}).

Let $(B,C)$ and $(B_V,C_V)$ be 
the sets of two- and one-form gauge fields 
of $\bZ_{N_c}\times \bZ_{N_f}$ group that arises in (\ref{bcf_group}).
They satisfy by definition
\begin{align}
 N_c B= d C \ ,
\quad
N_f  B_V= d C_V \ .
\label{BcBV}
\end{align}
\eqref{Ctransf} gives rise to
the gauge transformations of $B$ and $B_V$ 
as
\begin{equation}
B \to B +  d \Lambda\ , \quad
B_V \to B_V+ d  \Lambda_V \ .
\label{Btransf}
\end{equation}
The origin of the two-form gauge fields can be understood as follows.
We formally rewrite the field strengths $f$ and $F$ in terms of 
the gauge field $\hat{a}$ and $\hat{A}$:
\begin{equation}
 f = da -i a \wed a 
= 
d (\hat{a} - \fr{1}{N_c} C I_{N_c} )
-i
(\hat{a} - \fr{1}{N_c} C I_{N_c} )
\wed
(\hat{a} - \fr{1}{N_c} C I_{N_c} )
= 
\hat{f} - \fr{1}{N_c} d C I_{N_c},
\end{equation}
\begin{equation}
 F = d A -i A \wed A = d \hat{A} -i \hat{A} \wed \hat{A} 
-\fr{1}{N_{f}} d C_V I_{N_f}
 = \hat{F}  -\fr{1}{N_{f}} d C_V I_{N_f}.
\end{equation}
By noting that \er{Ctransf} acts on
the field strength $\hat{f}$ and $\hat{F}$ as
\begin{equation}
 \hat{f} \to \hat{f} + d\Lambda I_{N_c}\ ,
\quad
 \hat{F} \to \hat{F} + d\Lambda I_{N_f}\ ,
\end{equation}
respectively, 
$\hat{f} - \fr{1}{N_c} dC I_{N_c}$ 
and $\hat{F} - \fr{1}{N_f} dC_V I_{N_f}$ 
are interpreted as St\"uckelberg couplings.
Therefore,
$\fr{1}{N_c} dC $ and $ \fr{1}{N_f} dC_V I_{N_f}$ are
naturally regarded as 
two-form gauge fields 
with the transformation law \er{Btransf}.
The $\bb{Z}_{N_c}\times\bb{Z}_{N_f}$ gauge fields 
are characterized by the vanishing field strength and 
non-trivial  Aharonov-Bohm phase.
In fact, 
the field strength of the two-form $B$ vanishes locally, 
$d B =\fr{1}{N_c} d d C =0$
and $B$ has $\bb{Z}_{N_c}$-valued Aharonov-Bohm phase,
$\int_{\cal S} B = \fr{1}{N_c}\int_{\cal S}dC \in \fr{2\pi}{N_c} \bb{Z} $.
Similar arguments hold for $B_V$.

Now we derive the BCF anomaly \cite{anbpop1,anbpop2}.
This is obtained from the $U(1)_A$ anomaly in
\eqref{dGdY} by replacing
\begin{align}
  \tr (f\wedge f)& \to 
  \tr ((\hat{f} - B) \wedge (\hat{f} - B))
 = 
\tr (\hat{f}\wedge \hat{f})-N_c\,B\wedge B
\\
  \tr (F\wedge F)&\to 
  \tr ((\hat{F} - B_V) \wedge (\hat{F} - B_V))
=
\tr (\hat{F}\wedge \hat{F})-N_f\,B_V\wedge B_V
\\
F_B& \to \hat{F}_B-nB-B_V \ ,
\end{align}
where $\tr \hat{f} = dC = N_c B$
and $\tr \hat{F} = dC_V = N_f B_V$.
It is found that
\begin{align}
\delta_A(\Gamma+\cY)
=
&-2\pi N_c n_A\int \frac{1}{8\pi^2}B\wedge B
\nn
\\
&
+\frac{2\pi \,d(R) n_A}{2N_fT(R)}
\int\frac{1}{8\pi^2}
\left(
\tr (\hat{F}\wedge\hat{F})-N_fB_V\wedge B_V
\right)
+\frac{2\pi \,d(R) n_A}{2T(R)}
\int\frac{1}{8\pi^2}
\left(\hat{F}_B-nB-B_V\right)^2
 \ ,
\label{bcf:anomaly}
\end{align}
mod $2\pi\bZ$.
The first term is a manifestation of the mixed 't Hooft anomaly between
the axial $\bZ_{2N_fT(R)}=\bZ_{4N_fT(R)}/\bZ_2$ and the one-form
$\bZ_{N_c}$ symmetry.

It is possible to remove the first term
by adding the local counter term
\begin{align}
 & \cY_B=
\frac{4N_fT(R)}{2\pi}(1+kN_c)\int
\chi\wedge\left(
C_3-\frac{1}{4\pi}B\wedge C
\right)\ ,
\label{C3counter}
\end{align}
with $k\in\bZ$.
Here, $C_3$ is a background three-form gauge field
with a gauge transformation by a two-form gauge parameter $\Lambda_2$,
\begin{equation}
 C_3 \to C_3 + d\Lambda_2. 
\label{C3transf}
\end{equation}
The three-form gauge field is normalized by the flux quantization 
condition,
\begin{align}
  \int dC_3 \in 2\pi\bZ \ .
\label{intdC3}
\end{align}
Further,
$C_3$ is defined to make a GS transformation
\begin{align}
  C_3\to C_3+\frac{N_c}{4\pi}\Big(
2B\wedge\Lambda+\Lambda\wedge d\Lambda\Big) \ ,
\label{GS}
\end{align}
under the one-form $\bZ_{N_c}$ gauge transformation 
\begin{align}
  B\to B+d\Lambda \ ,~~~C\to C+N_c\Lambda \ .
\label{gtr:ZNc}
\end{align}
The field strength of $C_3$ can be defined as
\begin{equation}
 G_4 : = dC_3 - \fr{N_c}{4\pi} B \wed B
\label{G4}
\end{equation}
This guarantees that $\cY_B$ is left invariant under
the one-form $\bZ_{N_c}$ gauge transformation.
$C_3$ plays an essential role in exploring the higher-group
structure in the low energy effective theory in the
QCD-like theories.
Hereafter, we set $k=0$ for simplicity,
since the term proportional to $k$ does not contribute to 
cancel the anomaly.

\section{BCF anomaly from low energy effective action}

In this subsection, we discuss how the BCF anomaly is reproduced by the 
low energy effective theories of pions in
the QCD-like theories.

\subsection{Vacuum structure and construction of chiral Lagrangian}

We assume that the QCD-like theories develop a fermion bilinear
condensate. Then, the low energy effective theory
is described by a nonlinear $\sigma$-model constructed from nonlinear
realization of the spontaneous chiral symmetry breaking. We refer to
it as a chiral Lagrangian with the degrees of freedom 
given by the pions.

When $R$ is complex, the order parameter reads
\begin{align}
\begin{array}{ccccc}
&SU(N_f)_L &SU(N_f)_R &U(1)_V &U(1)_A  \\
  \psi_{L\alpha}^{ai}\bar{\psi}_{Rai^{\prime}}^{\alpha}
& \Box&\bar{\Box}&0&2 
\end{array}
\end{align}
The bilinear condensate 
\begin{align}
\langle \psi_{L}\bar{\psi}_{R}\rangle=\Lambda^3
I_{N_f} \ ,
\label{ref:Rcomp}
\end{align}
gives rise to spontaneous
symmetry breaking for the nonabelian chiral symmetry
\begin{align}
  SU(N_f)_L\times SU(N_f)_R \to SU(N_f)_V \ .
\label{SULRV}
\end{align}
{}Furthermore, the bilinear condensate breaks
the discrete axial symmetry as
\begin{align}
  \bZ_{4N_fT(R)}\to \bZ_2 \ ,
\label{ZtoZ}
\end{align}
where $\bZ_2$ acts as $\psi_{L,R}\to -\psi_{L,R}$. 
Then, the vacuum manifold due to the bilinear condensate
is given by
\begin{align}
 \frac{\bZ_{4N_fT(R)}/\bZ_2\times (SU(N_f)_L\times SU(N_f)_R)/SU(N_f)_V}
{\bZ_{N_f}} \ .
\label{vmfd:Rcomp}
\end{align}
Here, $\bZ_{N_f}$ is the center of either $SU(N_f)_L$ or $SU(N_f)_R$,
which identifies two distinct phases of the condensate that are 
generated by the $\bZ_{2N_fT(R)}=\bZ_{4N_fT(R)}/\bZ_2$ action.
The low energy dynamics of the corresponding Nambu-Goldstone(NG)
bosons are described by a chiral Lagrangian
\begin{align}
  \cL=-\frac{f_\pi^2}{4}
\tr (\hU^{\dagger}\partial_\mu \hU)^2
\ .
\end{align}
Here,
\begin{align}
  \hU(\pi)=\exp\left(\frac{2i\pi}{f_\pi}\right)\in U(N_f) \ ,
\end{align}
is the exponentiated pion field, which transforms as
\begin{align}
  U(N_f)_L\times U(N_f)_R:~
\hU\to g_L \hU g_R^\dagger \ .
\end{align}
Note that the $U(1)$ part of $\hU$, the $\eta^\prime$ meson, is not 
massless 
because it is associated with 
the spontaneously breaking of the discrete symmetry
$\bZ_{2N_fT(R)}\subset U(1)_A$.
This fact is understood by a dynamically generated potential
for the $\eta^\prime$ meson as discussed shortly.

In the presence of the background gauge fields for 
the gauge group (\ref{bcf_group}) as well as 
$\bZ_{4N_fT(R)}\subset U(1)_A$,
the derivatives in the chiral Lagrangian must be written
in terms of the covariant derivative
defined by
\begin{align}
D\hU
=d\hU-i\Ab_L\hU+i\hU\Ab_R \ ,
\label{DU:Rc}
\end{align}
where $\Ab_{L,R}$ is defined in (\ref{AbLR}) and rewritten in terms
of the uplifted gauge fields as
\begin{align}
\Ab_L&=\Ab-\chi I_{N_f}=\hA-\left(\hA_B-\frac{n}{N_c}C+\chi\right)I_{N_f} 
\ ,
\nn\\
\Ab_R&=\Ab+\chi I_{N_f}=\hA-\left(\hA_B-\frac{n}{N_c}C-\chi\right)I_{N_f} 
\ .
\label{AbLRhat}
\end{align}

When $R$ is real, 
a gauge invariant fermion bilinear is given by
\begin{align}
  \epsilon^{\alpha\beta}\,\Psi_\alpha^{aI}\Psi_\beta^{aJ} \ .
\end{align}
This is symmetric under the exchange $I\leftrightarrow J$.
We assume that the QCD-like theory exhibits the nonabelian chiral 
symmetry breaking 
\begin{align}
  SU(2N_f) \to SO(2N_f) \ ,
\label{SUSO}
\end{align}
due to the vacuum condensate
\begin{align}
  \langle \epsilon^{\alpha\beta}\,\Psi_\alpha^{aI}\Psi_\beta^{aJ} \rangle
=\Lambda^3 \delta^{IJ} \ .
\label{ref:Rr}
\end{align}
As a consistency check for this assumption, this is in accord with
the Vafa-Witten theorem \cite{VW}, which
states that vector-like global symmetry is unbroken for any
vector-like gauge theory with the vacuum angle $\theta=0$.
We note that $SU(N_f)\subset SO(2N_f)$ 
is identified with $SU(N_f)_V$.
{}For recent studies of the chiral flavor symmetry breaking for
the adjoint QCD, see \cite{cordum} for instance.

We find that the bilinear condensate (\ref{ref:Rr})
leads to the vacuum manifold 
\begin{align}
 \frac{\bZ_{4N_fT(R)}/\bZ_2\times SU(2N_f)/SO(2N_f)}
{\bZ_{2N_f}} \ ,
\label{vmfd:Rr}
\end{align}
with $\bZ_{2N_f}$ being the center of $SU(2N_f)$.
The low energy effective is given by the chiral Lagrangian associated
with the spontaneous breaking $U(2N_f)\to SO(2N_f)$
\begin{align}
  \cL=-\frac{f_\pi^2}{4}
\tr (\hU^{\dagger}\partial_\mu \hU)^2
\ .
\end{align}
Here, $f_\pi$ is the decay constant for the QCD-like theory and
\begin{align}
  \hU(\pi)
\in
U(2N_f)/SO(2N_f) \ .
\label{hU:Rreal}
\end{align}
We utilize the same notation of the decay constant and the
exponentiated pion field as in the case of $R=\,$complex. 
$\hU$ transforms under $U(2N_f)$ as
\begin{align}
  U(2N_f):~
\hU\to g\, \hU g^T \ .
\end{align}
{}For a construction of this action using the Callan-Coleman-Wess-Zumino
(CCWZ) procedure,
see \cite{CCWZ}.
The $U(1)$ pion $\pi^0$ (referred to as the $\eta^\prime$ meson too) 
is massive 
because it is associated with 
the spontaneously broken chiral symmetry
$\bZ_{2N_fT(R)}\subset U(1)_A$ as shown in (\ref{vmfd:Rr}).
This fact is accounted for by a dynamically generated potential
for the $\eta^\prime$ meson.

Let $\cAb$ be the background gauge field for $U(2N_f)$.
In the presence of it, the covariant derivative of $\hU$ reads
\begin{align}
  D\hU=d\hU-i\cAb\hU-i\hU\cAb^T \ .
\label{DU:Rr}
\end{align}
$\Ab_L$ and $\Ab_R$, 
the background gauge fields for the $U(N_f)_L\times U(N_f)_R$ subgroup,
are embedded into $\cAb$ as
\begin{align}
  \cAb=\left(
\begin{array}{cc}
\Ab_L & 0   
\\
0 & -\Ab_R^{\ast}
\end{array}
\right) \ .
\label{cAALR}
\end{align}

Let us discuss the potential term of the $\eta^\prime$ meson
for a generic representation $R$, which makes the $\eta^\prime$ meson
massive.
We note that the $U(1)_A$ symmetry acts on the exponentiated pion field as
\begin{align}
  \hU \to e^{-2i\alpha_A}\hU \ ,
\label{hU:u1A}
\end{align}
where $\alpha_A$ is defined in (\ref{def:alphaA}).
The $U(1)$ pion is extracted from $\hU$ as
\begin{align}
  \hU=e^{2i\pi/f_{\pi}}
=e^{2i\pi^{\ha}T^{\ha}/f_{\pi}}
=e^{2i\eta^\prime T^0/f_{\pi}}e^{2i\pi^{a}T^{a}/f_{\pi}} \ ,
\end{align}
with
\begin{align}
  a=\left\{
\begin{array}{ll}
  1,2,\cdots,{\rm dim}\,SU(N_f) \ , & (R=\mbox{complex}) \\
  1,2,\cdots,{\rm dim}\,SU(2N_f)/SO(2N_f) \ . & (R=\mbox{real})
\end{array}
\right. 
\end{align}
Here, the generators $T^{\ha}$ are normalized as
\begin{align}
  \tr (T^{\ha}T^{\hb})=\frac{1}{2}\delta^{\ha\hb} \ ,
\end{align}
with the $U(1)$ generator given by
\begin{align}
  T^0=\left\{
\begin{array}{ll}
  I_{N_f}/\sqrt{2N_f} \ , & (R=\mbox{complex})\\
  I_{2N_f/}/\sqrt{4N_f}\ .& (R=\mbox{real})
\end{array}
\right. \ .
\end{align}
With this choice, $\eta^\prime$ is a canonically normalized real scalar.
It is also useful to rewrite the $U(1)$ meson as
\begin{align}
  \hU=e^{i\wt\eta}e^{2i\pi^{a}T^{a}/f_{\pi}}
\equiv e^{i\wt\eta}\,U\ ,
\label{UhatU}
\end{align}
which gives
\begin{align}
  \wt\eta=
\left\{
\begin{array}{ll}
  \sqrt{\frac{2}{N_f}}\,\frac{\eta^\prime}{f_{\pi}} \ , & (R=\mbox{complex})\\
  \sqrt{\frac{1}{N_f}}\,\frac{\eta^\prime}{f_{\pi}} \ .& (R=\mbox{real})
\end{array}
\right. 
\end{align}
(\ref{hU:u1A}) shows that $U(1)_A$ acts as a shift of $\eta$
\begin{align}
  \wt\eta\to\wt\eta-2\alpha_A \ .
\end{align}
We also note that
the anomalous $U(1)_A$ transformation induces the shift of the
vacuum angle 
\begin{align}
\theta\to\theta+4N_fT(R)\alpha_A \ .
\end{align}
Then, the potential term for $\wt\eta$, if exists, is required to depend on
a single variable $\theta+2N_fT(R)\wt\eta$, a $U(1)_A$ invariant
\begin{align}
 V(\theta+2N_fT(R)\wt\eta\,) \ .
\label{pot:eta}
\end{align}

This potential is consistent with the Witten-Veneziano 
formula \cite{Witten:1979vv,Veneziano:1979ec} 
for the $\eta^\prime$ meson mass for large $N_c$ QCD with $R=\Box$.
The potential energy is of $\cO(N_c^2)$ because it 
is dominated by the gluon loop effects.
The large $N_c$ limit should be defined
by regarding the vacuum angle as of $\cO(N_c)$. This
is explained by noting the gauge theory action
\begin{align}
  S=\cdots+\theta\int\frac{1}{8\pi^2} \tr f\wedge f
=N_c\left[
\cdots
+\frac{\theta}{N_c}\int\frac{1}{8\pi^2} \tr f\wedge f 
\right]\ ,
\end{align}
with $\theta/N_c$ held finite.
We also note that $\wt\eta$ should be
of $\cO(N_c^0)$ because the chiral Lagrangian takes the form
\begin{align}
  -\frac{f_{\pi}^2}{4}\tr \left(\hU^{\dagger}\partial_\mu\hU\right)^2
=
  \frac{f_{\pi}^2}{4}\left[
N_f(\partial_\mu\wt\eta)^2
-\tr \left(U^{\dagger}\partial_\mu U\right)^2\right] \ ,
\end{align}
with $f_{\pi}^2=\cO(N_c)$. 
It then follows that the vacuum energy density as a function of $\theta$
and $\wt\eta$ takes the form
\begin{align}
  V=N_c^2\,h\left(
\frac{\theta+2N_fT(R)\wt\eta}{N_c}\right) \ .
\label{V:largeN}
\end{align}
The periodicity of the vacuum angle is manifested
with a multi-branched function \cite{Witten:1980sp}.
{}For a generic value of $\theta$ of $\cO(N_c^0)$, which is consistent
with the large $N_c$ limit, $h$ can be approximated
with a quadratic function for each branch because
\begin{align}
  \frac{\theta+2N_fT(R)\wt\eta}{N_c}=\cO(N_c^{-1}) \ .
\end{align}
Then, we obtain
\begin{align}
V=\frac{1}{2}\,\chi_g\,\min_{k\in\bZ}\left( \theta+2\pi k
+2N_fT(R)\wt\eta\right)^2+\cO(N_c^{-1}) \ .
\end{align}
Here, $\chi_g$ is the topological susceptibility, being
of $\cO(N_c^0)$.
Rewriting it in terms of $\eta^\prime$, the canonically normalized
field, the mass squared of the $\eta^\prime$ meson is given by
\begin{align}
  m_{\eta^\prime}^2=\frac{2N_f}{f_{\pi}^2}\,\chi_g \ .
\end{align}
This is the Witten-Veneziano formula \cite{Witten:1979vv,Veneziano:1979ec}.
{}For a derivation of it using a holographic dual, 
see \cite{SS1}.

When $R={\tiny \yng(2), \,\yng(1,1)}$ and {\bf adj.}, 
the vacuum energy density takes the same form as in (\ref{V:largeN}).
A distinction from the case of $R=\Box$ is that
the quark field makes the $\cO(N_c^2)$ contribution to $V$, which
is comparable with the planar effects of the gluon.
We have to note that for a generic $\theta$
\begin{align}
  \frac{\theta+2N_fT(R)\wt\eta}{N_c}=\cO(N_c^{0}) \ ,
\end{align}
because
\begin{align}
  T(R)=\cO(N_c) \ .
\end{align}
This shows that no approximation of $h$ is allowed, implying that
no multi-branch structure of the potential is there.
The $\eta^\prime$ mass is of
$\cO(N_c^0)$, and therefore not suppressed in the large $N_c$ limit.

\subsection{WZW term and $\eta^\prime$ term}

The 't Hooft anomaly matching condition states that the low energy dynamics
of the QCD-like theories must
reproduce not only the BCF anomaly (\ref{bcf:anomaly})
but also the full nonabelian anomaly that is obtained for
a generic rotation angle $\hat{\alpha}$ defined in (\ref{halpha}).
The anomaly matching condition for the BCF anomaly only is not
powerful enough to determine the low energy effective theory uniquely,
because it is a discrete anomaly. 
The 't Hooft anomaly matching for the nonabelian 
flavor symmetry requires that the WZW term
be added to the low energy effective theory.
With this action, the anomaly 
matching for the second and the third terms of the BCF anomaly
is satisfied.
It is argued that the first term of the BCF anomaly is reproduced
by another interaction term of the $\eta^\prime$ meson.
{}For a recent analysis of WZW terms in the presence
of background fields for discrete one-form symmetries, see also
\cite{Anber:2021iip}

We first discuss the cases for $R$ complex.
The WZW term reads \cite{Witten:1983tw}
\begin{align}
S_{\rm WZW}
=-i\frac{d(R)}{48\pi^2}\int_{M^4}Z
-i\frac{d(R)}{240\pi^2}\int_{M^4\times \bR}\tr(\hU d\hU^{-1})^5 \ ,
\label{WZW}
\end{align}
with
\begin{align}
  Z=&~
i\,\tr[(\Ab_R d\Ab_R+d\Ab_R \Ab_R-i\Ab_R^3)(\hU^{-1}\Ab_L\hU+i\hU^{-1}d\hU)-{\rm{p.c.}}]\nn\\
&-\tr[ d\Ab_Rd\hU^{-1}\Ab_L \hU-{\rm{p.c.}}]
-i\,\tr[\Ab_R(d\hU^{-1}\hU)^3-{\rm{p.c.}}]\nn\\
&-\frac{1}{2}\tr[(\Ab_Rd\hU^{-1}\hU)^2-{\rm{p.c.}}]
-\tr[\hU\Ab_R \hU^{-1}\Ab_L d\hU d\hU^{-1}-{\rm{p.c.}}]\nn\\
&-i\,\tr[\Ab_R d\hU^{-1}\hU\Ab_R \hU^{-1}\Ab_L\hU-{\rm{p.c.}}]
+\frac{1}{2}\tr[(\Ab_R\hU^{-1}\Ab_L\hU)^2]\ .
\label{Z}
\end{align}
Here, ``p.c.''~represents the terms obtained by
making the exchange $\Ab_L\leftrightarrow \Ab_R$ and 
$\hU\leftrightarrow \hU^{-1}$.
As well-known, this yields the chiral anomaly 
of L-R form for a generic $\Ab_L$ and $\Ab_R$:
\begin{align}
  \delta S_{\rm WZW}
=
-\frac{d(R)}{24\pi^2}\int\tr 
\left[
\alpha_L\,d\left(
\Ab_Ld\Ab_L-\frac{i}{2}\Ab_L^3\right)
-
\alpha_R\,d\left(
\Ab_Rd\Ab_R-\frac{i}{2}\Ab_R^3\right)
\right] \ ,
\label{dSWZW}
\end{align}
with 
$g_L=e^{i\alpha_L},~g_R=e^{i\alpha_R}$.
Inserting $\hU=e^{i\wt\eta}\,U$ and (\ref{AbLRhat}) into
the WZW term gives
\begin{align}
  S_{\rm WZW}=S_{\rm WZW}^{(0)}
&+i\frac{d(R)}{48\pi^2}\int (d\wt\eta+2\chi)
\,\tr \left[\Fb\left(U^{-1}DU+UDU^{-1}\right)\right]
\nn\\
&
+\frac{d(R)}{48\pi^2}\int d\wt\eta\,
\tr \Big(
6\Ab d\Ab-4i\Ab^3
\Big) 
+\frac{d(R)}{48\pi^2}\int \chi\,
\tr \Big(
8\Ab d\Ab-6i\Ab^3
\Big) \ .
\end{align}
Here,
\begin{align}
  S_{\rm WZW}^{(0)}=S_{\rm WZW}\big|_{\wt\eta=\chi=0} \ ,
\end{align}
and
\begin{align}
  DU=dU-iA_LU+iUA_R \ ,~~~
  DU^{-1}=dU^{-1}-iA_RU^{-1}+iU^{-1}A_L \ ,
\nn
\end{align}
with $A_{L,R}=\Ab+A_BI_{N_f}$ being the background gauge
fields for $SU(N_f)_L$ and $SU(N_f)_R$, respectively.
It is found that the $U(1)_A$ gauge transformation
\begin{align}
  \delta_A\wt\eta=-2\alpha_A \ ,~~
  \delta_A\chi=d\alpha_A \ ,
\end{align}
with $(\alpha_R-\alpha_L)/2=\alpha_AI_{N_f}$ 
leads to 
\begin{align}
  \delta_A S_{\rm WZW}
=
-\frac{d(R)}{12\pi^2}\int\alpha_A\,\tr 
\left(
\Fb^2-\frac{i}{2}\Fb\Ab^2
\right)
\end{align}
The local counter term $\cY_f$ given in (\ref{cYcf})
is written only by the background gauge fields so that
it remains to be a local counter term in the IR as well.
With this term added, the variation of the WZW action is changed as
\begin{align}
  \delta_A(S_{\rm WZW}+\cY_f)
&=
\frac{d(R)}{4\pi^2}\int \alpha_A\,\tr (\Fb\wedge\Fb)
\nn\\
&=
\frac{d(R)}{4\pi^2}\int
\alpha_A\left[ {\rm tr}(\hF-B_V)^2
+N_f\left(\hF_B-nB-B_V\right)^2
\right] \ ,
\label{dSY}
\end{align}
reproducing the second and the third term of the BCF anomaly in 
(\ref{bcf:anomaly}) upon setting $\alpha_{A}=2\pi n_A/4N_fT(R)$.

We note that, in the presence of the background gauge 
fields $B$ and $B_V$,
$S_{\rm WZW}$
suffers an operator-valued ambiguity of how to extend it to a five-dimensional
local action. 
To see this, suppose that $S_{\rm WZW}$ is defined on  
five-manifolds 
$X_1$ and $X_2$ with a common boundary equal to
the four-manifold where the QCD-like theory is defined. 
The difference of the actions is obtained by evaluating
the action on the compact manifold $X_1\cup \bar{X}_2$:
\begin{align}
&-\frac{d(R)}{8\pi^2}\int_{X_1\cup \bar{X}_2} d\wt\eta\wedge
\left(\tr \Fb^2
+\frac{i}{6}\,d\,\tr \!\left[\Fb\left(U^{-1}DU+UDU^{-1}\right)\right]
\right) 
=
-\frac{d(R)}{8\pi^2}\int_{X_1\cup \bar{X}_2} d\wt\eta\wedge
\tr \Fb^2
\,\notin\, 2\pi\bZ \ .
\label{qamb:wzw}
\end{align}
The integral of 
$d\,\tr \!\left[\Fb\left(U^{-1}DU+UDU^{-1}\right)\right]$
is trivial because $\tr \!\left[\Fb\left(U^{-1}DU+UDU^{-1}\right)\right]$
is gauge invariant.
As will be shown, this ambiguity is eliminated automatically
by adding an interaction term between the $\eta^{\prime}$ meson
and the background $B$ in such a way that
the first term of the BCF anomaly (\ref{bcf:anomaly})
can be reproduced in the low energy effective action.

The WZW term for $R$ real is derived in the appendix
\ref{leet:Rr}.
It exhibits the same operator-valued ambiguity (\ref{qamb:wzw}) as in the case
for $R$ complex. This is easy to verify by
inserting (\ref{cAALR}) and (\ref{AbLRhat}) into the WZW term (\ref{Zr2}).
This ambiguity can be cancelled as well by requiring the anomaly
matching of the full BCF anomaly.

Now that the WZW term is found to reproduce 
the second and the third terms of
the BCF anomaly given in (\ref{bcf:anomaly}), we discuss how 
the first term of the BCF anomaly is matched
in terms of the effective action of the pions.
We argue that the anomaly matching is achieved 
by adding to the effective action
\footnote{The local counterterm composed of $\eta^\prime$
and $B\wedge B$ is considered also in \cite{Kitano:2020evx}.}
\begin{align}
S_{\eta BB}= 
-\frac{2N_fT(R)}{2\pi}(1+\ell N_c)\int \wt\eta
\left(dC_3-\frac{N_c}{4\pi}B\wedge B \right)
 \ .
\label{etaBB}
\end{align}
Here, $\ell$ is an integer that is not fixed from 
the discrete anomaly matching condition we are employing.
$C_3$ is the background three-form gauge potential, which 
is defined in \er{C3counter}.
To see why, we first note that 
the discrete axial symmetry $\bZ_{2N_fT(R)}=\bZ_{4N_fT(R)}/\bZ_2$
acts as
\begin{align}
\bZ_{2 N_fT(R)}:~  \wt\eta\to\wt\eta-\frac{2\pi}{2N_fT(R)} \ .
\label{shift:eta}
\end{align}
This shift reproduces the mixed 't Hooft 
anomaly between the $\bZ_{2N_fT(R)}$ and the one-form $\bZ_{N_c}$
symmetry modulo $2\pi\bZ$, because
$C_3$ gives no contribution to the 't Hooft anomaly 
thanks to the normalization condition (\ref{intdC3}).
{}Furthermore, 
$S_{\eta BB}$ is 
left invariant under the one-form $\bZ_{N_c}$ gauge transformation
(\ref{gtr:ZNc})
together with the GS-type transformation law (\ref{GS}).
Hereafter, we set $\ell=0$ for simplicity.

We remark that instead of (\ref{etaBB}), the mixed 't Hooft anomaly 
may be obtained from an interaction term
between $\wt\eta$ and the gluon, 
\begin{equation}
 S_{\eta ff}
 = - \fr{2 N_f T(R)}{8\pi^2} \int \wt\eta\, 
\tr ((\hat{f} - B) \wed (\hat{f} - B)) \ . 
\end{equation}
This term is studied in \cite{DiVecchia:1980yfw,DiVecchia:2017xpu} to 
derive the Witten-Veneziano formula for the
$\eta^\prime$ mass.
In this scenario, 
the background gauge field $C_3$ plays no role in achieving the
anomaly matching condition. Instead, it is assumed
that the gluon fields are confined in the low energy regime.
$C_3$ is interpreted as the background gauge field of an 
emergent two-form symmetry whose 
charged object is an $\eta^\prime$ vortex.
To see this, we note that $\eta^\prime$ can have a winding number,
\begin{equation}
 \int_{\cal C} d\wt\eta \in 2\pi \bb{Z}\ ,
\label{etawind}
\end{equation}
since it may be understood as a (pseudo-)NG boson of 
the $U(1)_A$ symmetry.
The integrand $d\wt\eta $ 
obeys the analog of the 
Bianchi identity,
\begin{equation}
 dd \wt\eta =0\ .
\end{equation}
Therefore, the integral \er{etawind} is topological 
under a small deformation 
${\cal C} \to {\cal C} \cup \der {\cal S}_0$,
\begin{equation}
 \int_{{\cal C} \cup \der {\cal S}_0} d\wt\eta 
-
 \int_{{\cal C}} d\wt\eta
 = 
 \int_{ \der {\cal S}_0} d\wt\eta
= 
 \int_{{\cal S}_0} d d\wt\eta 
= 0\ ,
\end{equation} 
where ${\cal S}_0$ is a surface with a boundary.
We thus have a unitary topological object parameterized by 
$e^{i\gamma} \in U(1)$,
\begin{equation}
 U_2 ({\cal C}, e^{i\gamma})   
=
 e^{\fr{i\gamma}{2\pi} \int_{\cal C} d\wt\eta }\ .
\end{equation}
The topological object is regarded as a
symmetry generator for a $U(1)$ two-form symmetry, 
because
the generator acts on the vortex string operator 
$V_\eta ({\cal S})$ with the worldsheet ${\cal S}$ as
\begin{equation}
  U_2 ({\cal C}, e^{i\beta}) V_\eta ({\cal S})
 = e^{i\gamma \link ({\cal S,C})} V_\eta ({\cal S})\ ,
\end{equation}
where $\link ({\cal S,C})$ denotes the linking number between 
${\cal S}$ and ${\cal C}$.
The conserved current for the two-form symmetry is identified with
$\fr{1}{2\pi} d\wt\eta$ and couples minimally to $C_3$.

\subsection{Cancellation of the operator-valued ambiguity}

$S_{\eta BB}$ is written
in the form of the integral of a local functional 
on a five-manifold with the boundary
\begin{align}
  S_{\eta BB}=
-\frac{2N_fT(R)}{2\pi}\int_{X_1} d\wt\eta\wedge
\left(dC_3-\frac{N_c}{4\pi}B\wedge B \right)
 \ .
\end{align}
This depends on how the action is extended to five dimensions
because
\begin{align}
-2N_fT(R)\int_{X_1\cup\bar{X}_2} \frac{N_c}{8\pi^2}\,d\wt\eta\wedge
B\wedge B \,\notin\, 2\pi\bZ \ .
\label{qamb:BB}
\end{align}
This is the manifestation of another operator-valued ambiguity
in addition to that from the WZW term.

The net operator-valued ambiguity associated with the $\eta^{\prime}$ meson
reads
\begin{align}
  \int_{X_1\cup\bar{X}_2} d\wt\eta\wedge
\left[
-2N_fT(R)\,\frac{N_c}{8\pi^2}B\wedge B+
\frac{d(R)}{8\pi^2}\,\tr \Fb^2
\right] \ .
\end{align}
Cancellation of the operator-valued ambiguity requires that
this take values in $2\pi\bZ$.
As $\wt\eta$ is a $2\pi$-periodic boson, this condition
is stated
in terms of the background gauge fields only
\begin{align}
  \int_{X} 
\left[
-2N_fT(R)\,\frac{N_c}{8\pi^2}B\wedge B+
\frac{d(R)}{8\pi^2}\,\tr \Fb^2
\right]\,\in\, \bZ \ .
\end{align}
$X$ is any compact spin four-manifold.
This integral is always integer because 
this is equal to the index of the Dirac operator (mod $\bb{Z}$)
whose 
gauge field is $\wt{A}$ satisfying the proper normalization.

This can be verified  
by taking $X=T^2\times T^2$ and
turning on the 't Hooft fluxes on it, whose explicit form is
given in \cite{anbpop1}.
Let $(x^1,x^2)$ and $(x^3,x^4)$ be the coordinates of the two two-tori
with $x^{1,2,3,4}\sim x^{1,2,3,4}+2\pi$.
Then,
\begin{align}
  B=\frac{1}{2\pi N_c}
\left(
m_{12}\,dx^1\wedge dx^2
+m_{34}\,dx^3\wedge dx^4\right)
\ .~~~(m_{12},m_{34}\in\bZ\mbox{ mod }N_c)
\end{align}
\begin{align}
  \hF&=\frac{1}{2\pi}
\left(
m_{12}^f\,dx^1\wedge dx^2
+m_{34}^f\,dx^3\wedge dx^4\right)
\left(
\begin{array}{cccc}
1 & &&\\
& 0 && \\
&& \ddots &  \\
&&& 0
\end{array}
\right) \ ,
\nn\\
B_V&=\frac{1}{2\pi N_f}
\left(
m_{12}^f\,dx^1\wedge dx^2
+m_{34}^f\,dx^3\wedge dx^4\right)
\ .~~~(m_{12}^f,m_{34}^f\in\bZ\mbox{ mod }N_f)
\end{align}
\begin{align}
F_B=\frac{1}{2\pi}
\left(
m_{12}^B\,dx^1\wedge dx^2
+m_{34}^B\,dx^3\wedge dx^4\right)
\ .~~~(m_{12}^B,m_{34}^B\in\bZ)
\end{align}
As a consistency check, we note $\hF-B_V\in\mathfrak{su}(N_f)$.
It is found that, modulo $\bZ$ 
\begin{align}
&  \int_{T^2\times T^2} 
\left[
-2N_fT(R)\,\frac{N_c}{8\pi^2}B\wedge B+
\frac{d(R)}{8\pi^2}\,\tr \Fb^2
\right]
\nn\\
=&\,-2N_fT(R)\frac{m_{12}m_{34}}{N_c}
+\frac{d(R)}{N_c}\left[
n(m_{12}^fm_{34}+m_{34}^fm_{12})
-nN_f(m_{12}^Bm_{34}+m_{34}^Bm_{12})
+\frac{n^2N_f}{N_c}m_{12}m_{34}
\right] \ .
\end{align}
It is easy to see that this is always an integer.

\subsection{Higher-group structure}

It is found that the low energy effective action of the QCD-like theory 
is given by
\begin{align}
  S_{\rm LEET}=&\int d^4x\left[-\frac{f_{\pi}^2}{4}
\tr \left(\hU^{\dagger} D_\mu\hU\right)^2
-V(\theta+2N_fT(R)\wt\eta\,)\right]
+S_{\rm WZW}+\cY_f+S_{\eta BB} \ .
\label{leet:bcf}
\end{align}
In this subsection, 
we specify the higher-group structure of the background gauge fields 
that is encoded in
(\ref{leet:bcf}) and show that
this is identified with a semistrict three-group
(two-crossed module).

The semistrict three-group is given by the following ingredients.
\begin{enumerate}
 \item A triple of groups, $(G, H, L)$.
\item Maps between groups, $\der_1: H \to G$, 
and $\der_2: L \to H$. 
They are homomorphism with respect to the group composition,
$\der_1 (h_1 h_2) = (\der_1 h_1) (\der_1 h_2)$, 
and $\der_1 (l_1 l_2) = (\der_1 l_1) (\der_1 l_2)$ 
for $h_1, h_2 \in H$ and $l_1, l_2 \in L$,
respectively.  
\item Action $\trr$ of $G$ on $G, H, L$:
$g \trr g' := gg' g^{-1} \in G$, $g \trr h \in H$, and $g \trr l \in L$.
This operation is a generalization of the adjoint transformation.
\item Peiffer lifting, $\{h_1,h_2\} \in L$ for $h_1, h_2 \in H$.
\item Consistency between the above operations, 
e.g., $g \trr \{h_1, h_2\} = \{g\trr h_1, g \trr h_2\}$.
\end{enumerate}
For more detail, see e.g., \cite{Hidaka:2020izy}.

{}From the above data, we can construct a gauge theory for the three-group,
which consists of zero-, one-, and two-form gauge field which are valued on 
the Lie algebra of $G$, $H$, and $L$, respectively.
For the Lie groups $G$, $H$, $L$, 
the Lie algebra of the three-group can be introduced as follows.
\begin{enumerate}
 \item A triple of Lie algebra $(\frak{g}, \frak{h}, \frak{l})$
of the triple of Lie groups $(G,H,L)$.
\item Maps between groups, $\der_1: \frak{h} \to \frak{g}$, 
and $\der_2: \frak{l} \to \frak{h}$. 
They are homomorphism with respect to the Lie bracket,
$\der_1 ([\ul{h}_1, \ul{h}_2])
 = [\der_1 \ul{h}_1, \der_1 \ul{h}_2]$, 
and 
$\der_2 ([\ul{l}_1, \ul{l}_2])
 = [\der_1 \ul{l}_1, \der_1 \ul{l}_2]$
for $\ul{h}_1, \ul{h}_2 \in \frak{h}$ and $\ul{l}_1, \ul{l}_2 \in \frak{l}$,
respectively.  
They are boundary maps satisfying $\der_1 \circ \der_2\, l = 1 \in G$ 
for all $l \in L$.
\item Action $\trr$ of $\frak{g}$ on $\frak{g,h,l}$:
$\ul{g} \trr \ul{g}' := [\ul{g},\ul{g'}] \in \frak{g}$, 
$\ul{g} \trr \ul{h} \in \frak{h}$, and 
$\ul{g} \trr \ul{l} \in \frak{l}$.
This operation is a generalization of the adjoint transformation.
\item Peiffer lifting, $\{\ul{h}_1,\ul{h}_2\} \in \frak{l}$ 
for $\ul{h}_1, \ul{h}_2 \in \frak{h}$.
\item Consistency between the above operations, 
e.g., 
$\ul{g} \trr \{\ul{h}_1, \ul{h}_2\} = \{\ul{g}\trr \ul{h}_1,\ul{g}\trr \ul{h}_2\}$.
\end{enumerate}
We then introduce the gauge fields for the three-group,
which consist of $\frak{g}$-, $\frak{h}$-, and $\frak{l}$-valued 
one-, two-, and three-form fields 
$A= A^\alpha u_\alpha$, $B = B^a v_a $, and $C = C^A w_A$, 
satisfying the following gauge transformation laws,
\footnote{In this subsection, the gauge fields are taken to be
anti-Hermitian.}
\begin{equation}
\begin{split}
 A &
\to 
A' = g \trr A + gdg^{-1} + \der_1 \ul{h},
\\
\quad
B 
\to 
& B' = g \trr B + d\ul{h} - \ul{h} \wed \ul{h}
+ A' \trr \ul{h} + \der_2 \ul{l},
\\ 
C
&
\to 
C' =
 g \trr C + d\ul{l} + A' \trr \ul{l}
 + \{\der_2 \ul{l} , \ul{h}\}
- 
\{B',\ul{h}\}
-
\{\ul{h}, g^{-1} \trr B\},
\end{split}
\label{3transf}
\end{equation}
 and field strengths,
\begin{equation}
 \begin{split}
  F  & = dA + A \wed A,
\\
H & = dB + A \trr B,
\\
G & = dC + A \trr C +\{B, B\}.
 \end{split}
\label{3fs}
\end{equation}
Here, $g$, $\ul{h} = \ul{h}^a v_a $, and $\ul{l} = \ul{l}^A w_A$ 
are $G$-, $\frak{h}$-, and $\frak{l}$-valued 
zero-, one-, and two-form parameters.
We have taken the bases of 
$\frak{g}$, $\frak{h}$, and $\frak{l}$ 
as $\{u_\alpha\}$, $\{v_a\}$, and $\{w_A\}$,
respectively.
The operations 
$\der_1$, $\der_2$, $\trr $, and $\{\}$ on the differential 
forms are given by 
the wedge products with the operations for the Lie algebra:
\begin{equation}
 g \trr A = A^\alpha g \trr u_\alpha 
= A^\alpha g u_\alpha g^{-1},
\end{equation}
\begin{equation}
 \der_1 \ul{h} = \ul{h}^a (\der_1 v_a),
\end{equation}
\begin{equation}
 \ul{h} \wed \ul{h}
= \fr{1}{2} \ul{h}^a \wed \ul{h}^b [v_a, v_b],
\end{equation}
\begin{equation}
\{\der_2 \ul{l}, \ul{h}\}
= \ul{l}^A \wed \ul{h}^a \{(\der_2 w_A), v_a\},
\end{equation}
and so on.
It is possible to take the following conditions,
\begin{equation}
  F - \der_1 B  =0 ,
\quad
H - \der_2 C  =0, 
\end{equation}
which are called fake curvature vanishing conditions.

Conversely, we can specify the three-group structure 
of the background gauge fields in an
appropriate class of a gauge theory with given one-, two-, and three-form 
gauge fields.
We first regard the group 
$SU(N_c)$ as a global symmetry group.
The background gauge fields $\hat{a}$, 
$\hat{A}$ and $\hat{A}_B$, $B$ and $B_V$, and $C_3$ 
are identified with those
for the gauge groups
\begin{equation}
 G_0 = U(N_c) \times U(N_f)_V \times U(1)_B
\cong \fr{SU(N_c) \times U(1)}{\bb{Z}_{N_c}} 
\times 
\fr{SU(N_f) \times U(1)}{\bb{Z}_{N_f}} 
\times U(1)_B,
\end{equation}
\begin{equation}
 H = \bb{Z}_{N_c} \times \bb{Z}_{N_f},
\end{equation}
\begin{equation}
 L = U(1).
\end{equation}
respectively.
For convenience, 
we write the elements of $G_0$ as 
\begin{equation}
 (g, e^{i\alpha_c}, g_V, e^{i\alpha_f}, e^{i\alpha_B}) 
\in SU(N_c) \times U(1) \times SU(N_f) \times U(1) \times U(1)_B,
\end{equation}
with the identifications by $\bb{Z}_{N_c}$ and $\bb{Z}_{N_f}$,
\begin{equation}
 (I_{N_c}, e^{2\pi i m_c/ N_c}, I_{N_f} ,1,1) 
\sim  (e^{2\pi i m_c/ N_c} I_{N_c}, 1, I_{N_f} ,1,1) ,
\label{UNcid}
\end{equation}
\begin{equation}
 (I_{N_c}, 1 , I_{N_f} , e^{2\pi i m_f/ N_f} ,1) 
\sim  (I_{N_c}, 1, e^{2\pi i m_f/ N_f} I_{N_f} ,1,1).
\end{equation}

We determine the structure of the three-group 
from the gauge transformation laws and field strengths.
Identification of the gauge transformation laws in \er{ahat_ij:fund}, \er{220124.1156} and \er{Btransf} with \er{3transf} shows that 
the map $\der_1: H \to G_0$ should read
\begin{equation}
\begin{split}
& \der_1 (e^{2\pi i m_c / N_c }, e^{2\pi i m_f / N_f })
\\
&
= ( I_{N_c},e^{2\pi i m_c / N_c }, I_{N_f},  e^{2\pi i m_f / N_f }  ,
e^{ - 2\pi i n m_c / N_c } e^{ - 2\pi i m_f / N_f }).
\end{split}
\label{mapHG}
\end{equation}
Meanwhile, the map $\der_2: L \to H$ is trivial,
\begin{equation}
 \der_2 (e^{i\gamma}) = (1, 1) \in \bb{Z}_{N_c} \times \bb{Z}_{N_f},
\end{equation}
for $e^{i\gamma } \in L$.
This is because the two-form gauge fields $ B$ and $B_V$ are
not transformed under the gauge transformation in \er{C3transf}.
We remark that the constraint 
$\tr (\hat{f}) = d C  = N_c B$ can be understood as the 
fake curvature vanishing condition imposed on the $U(1)$
sector of $U(N_c)$, i.e., 
$ 0 = \tr (\hat{f}) - \tr (\der_1 B) = \tr (\hat{f}) - \tr (B I_{N_c}) $.
Identification of the four-form field strength $G_4$ (\ref{G4}) 
with $G$ in \er{3fs}
leads to the Peiffer lifting, 
\begin{equation}
 \{ (e^{2\pi i m_c/ N_c} , e^{2\pi i m_f / N_f}),
  (e^{2\pi i m'_c /N_c } , e^{2\pi i m_f / N_f })\}
=
e^{- i \fr{N_c}{4\pi} \cdot \fr{2\pi m_c}{N_c} \cdot \fr{2\pi m'_c}{N_c} }
= e^{- i \pi m_c  m'_c / N_c  },
\end{equation}
where $(e^{2\pi i m_c /N_c } , e^{2\pi i m_f / N_f}) \in \bb{Z}_{N_c} \times \bb{Z}_{N_f}$.
\footnote{This definition of the Peiffer lifting has 
an ambiguity under $m_c \to m_c + N_c$.
In order to define the Peiffer lifting in an unambiguous way, 
we may need to include the spin structure of the spacetime. We leave this issue as future work.}

Now we promote the $SU(N_c)$ gauge field to dynamical degrees of freedom.
$SU(N_c)\subset G_0 $ should be treated as 
a gauge redundancy, i.e., the elements of $SU(N_c)$ in $G_0$ 
are regarded as the identity.
In particular, the gauging of $SU(N_c)$ gives rise to
further identifications in \er{UNcid}
and \er{mapHG}:
\begin{equation}
  (I_{N_c}, e^{2\pi i m_c/ N_c}, I_{N_f} ,1,1) 
\sim 
 (e^{2\pi i m_c/ N_c} I_{N_c}, 1, I_{N_f} ,1,1) 
\sim 
 (I_{N_c}, 1, I_{N_f} ,1,1) ,
\end{equation}
and 
\begin{equation}
\begin{split}
& \der_1 (e^{2\pi i m_c / N_c }, e^{2\pi i m_f / N_f })
\\
&
= ( I_{N_c},e^{2\pi i m_c / N_c }, I_{N_f},  e^{2\pi i m_f / N_f }  ,
e^{ - 2\pi i n m_c / N_c } e^{ - 2\pi i m_f / N_f })
\\
&
\sim 
 ( I_{N_c},1, I_{N_f},  e^{2\pi i m_f / N_f }  ,
e^{ - 2\pi i n m_c / N_c } e^{ - 2\pi i m_f / N_f }),
\end{split}
\label{mapHGgauge}
\end{equation}
respectively.
We find that the kernel of $\der_1$ is given by
\begin{equation}
 \ker \der_1 
= \{(e^{2\pi i m_c/N_c} ,1) \in \bb{Z}_{N_c} \times \bb{Z}_{N_f}| e^{2\pi i m_c /N_c} \in \bb{Z}_n \} \simeq \bb{Z}_{\gcd (N_c, n)}\ ,
\end{equation} 
ending up with a one-form symmetry group 
$\bb{Z}_{\gcd (N_c, n)}$.
This is identical to the subgroup of
$\bZ_{N_c}$ that 
acts nontrivially on
Wilson loop operators
in QCD-like theories~\cite{anbpop1}.

\subsection{Integrating out the $\eta^\prime$ meson}

We now focus on the low energy scales where the $\eta^\prime$ meson
is heavy enough to be integrated out.
Then, the potential (\ref{pot:eta}) 
gives rise to the spontaneous breaking of the discrete axial
symmetry $\bZ_{2N_fT(R)}$.
The low energy effective action for $\eta^\prime$
is given by replacing the kinetic term of $\wt\eta$ with
a BF action associated with the spontaneous $\bZ_{2N_fT(R)}$ breaking
\cite{kapsei}:
\begin{align}
  \frac{2N_fT(R)}{2\pi}\int \left(d\wt\eta+2\chi\right)\wedge c_3 \ ,
\end{align}
where $c_3$ is a dynamical three-form gauge field with
the normalization condition given by
\begin{align}
  \int dc_3\in 2\pi\bZ \ .
\end{align}
The resultant low energy effective action 
is given by
\begin{align}
  S=-\frac{f_\pi^2}{4}
\int d^4x
\,\tr \left(U^{\dagger} D_\mu U\right)^2 
+S_{\rm WZW}+\cY_f+S_{\rm BF} \ .
\label{S:noeta}
\end{align}
Here,
\begin{align}
 S_{\rm BF}&=
  \frac{2N_fT(R)}{2\pi}\int \left(d\wt\eta+2\chi\right)\wedge c_3
+S_{\eta BB}
\nn\\
&=\frac{2N_fT(R)}{2\pi}\int 
\left(d\wt\eta+2\chi\right)\wedge 
\left(c_3+C_3-\frac{1}{4\pi}B\wedge C
\right) -\cY_B \ .
\end{align}
It is useful to shift the dynamical gauge field as
$c_3\to c_3-C_3$, which keeps the normalization condition unchanged.
Then, the resultant $c_3$ makes a GS transformation
under the one-form $\bZ_{N_c}$ gauge symmetry transformation
\begin{align}
  c_3\to c_3+\frac{N_c}{4\pi}\left(2\Lambda\wedge B
+\Lambda\wedge d\Lambda\right) \ .
\end{align}

Summarizing the terms that involve the $\eta^{\prime}$ meson,
we find
\begin{align}
&  S_{\rm WZW}+\cY_f+S_{\rm BF}
\nn\\
=&-\frac{2N_fT(R)}{2\pi}\int\left(d\wt\eta+2\chi\right)
\left[
dc_3-\frac{N_c}{4\pi}B\wedge B
+\frac{d(R)}{8\pi N_fT(R)}\left\{
{\rm tr}(\Fb^2)
+\frac{i}{6}d\,{\rm tr}\left[\Fb\wedge\left(
U^{-1}DU+UDU^{-1}\right)\right]
\right\}\right]
\nn\\
&+\frac{2N_fT(R)}{\pi}\int\chi\wedge \left[
dC_3-\frac{N_c}{4\pi}B\wedge B
  +\frac{d(R)}{8\pi N_fT(R)}{\rm tr}\,\Fb^2
\right]+S_{\rm WZW}^{(0)}\ .
\label{BF:eta}
\end{align}
This reproduces the BCF anomaly evidently.
It is interesting to note that part of this action is equivalent
to the effective action of an axion that is constructed in
\cite{anbpop2}.

\section{Discussions}

In this paper, we have derived the low energy effective theories
of the QCD-like theories by requiring the 't Hooft anomaly
matching condition for the BCF anomaly.
This result is based on the assumption that the QCD-like theories
exhibit spontaneous chiral symmetry breaking due to the quark
bilinear condensate. Use of the BCF anomaly matching seems not to
be powerful enough to specify uniquely the phase structures for a given
$R$ and $N_f$. However, it might give us an important clue to a
classification of the phase structures by applying the methods
employed in this paper.
As an example, we assume instead that the QCD-like theories gives rise
to an exotic chiral symmetry breaking due to the condensate of
a gauge invariant operator other than the quark bilinear.\footnote{Such 
a scenario has been discussed recently in 
\cite{Yamaguchi:2018xse, Anber:2021lzb, Anber:2021iip} for instance.}
It would be interesting to examine if we can find the low energy
effective action of the Nambu-Goldstone bosons such that it satisfies 
the BCF anomaly
matching condition and furthermore any type of operator-valued ambiguity
disappears.

As pointed out before, the effective action (\ref{BF:eta})
is reminiscent of the axion effective action derived in
\cite{anbpop2}. It would be nice to clarify the relation 
of the two actions in more detail. 
{}For this purpose, we may couple the QCD-like theory with
a neutral and complex Higgs field as performed in \cite{anbpop2} with
a difference being that the Higgs coupling is tuned so that
the Higgs vev gives a light quark mass.
The light mass is proportional to an axion phase and generates a mass term 
of the pions.
It is expected that integrating out the massive pions leads to
the axion effective action.

\section*{Acknowledgements}

We would like to Erich Poppitz for a helpful discussion.
RY is supported by JSPS KAKENHI Grants No.~JP21J00480, JP21K13928.

\appendix

\section{Low energy effective action for $R$ real}
\label{leet:Rr}

We derive the low energy effective action of the QCD-like theory
for $R$ real in an assumption that it exhibits the quark bilinear
condensate.

\subsection{Chiral Lagrangian}
\label{lag:ccwz}
This chiral Lagrangian is constructed by using the 
Callan-Coleman-Wess-Zumino(CCWZ) procedure
\cite{CCWZ}. We first divide the generator of $SU(2N_f)$ into
the unbroken and broken ones respectively.
\begin{align}
  H^a&\in \mathfrak{so}(2N_f) \ ,~(a=1,2,\cdots\mbox{dim}\,SO(2N_f))
\nn\\
  X^\alpha&\in \mathfrak{u}(2N_f)-\mathfrak{so}(2N_f) \ ,
\end{align}
with the normalization condition given by
\begin{align}
  \tr (H^aH^b)=\frac{1}{2}\delta^{ab}  \ ,~~
  \tr (X^\alpha X^\beta)=\frac{1}{2}\delta^{\alpha\beta}  \ .
\end{align}

Since the coset $U(2N_f)/SO(2N_f)$ is a symmetric space, these obey
the commutation relations of the form
\begin{align}
  [H^a,H^b]=if^{abc}H^c \ ,~~~
[H^a,X^\alpha]=(t^a)^{\alpha\beta}X^\beta \ ,~~
[X^\alpha,X^\beta]=if^{\alpha\beta c}H^c \ .
\end{align}
Here, $t^a$ is the generators of $SO(2N_f)$ in the $\Box\!\Box$
representation. Consider an element of the coset $U(2N_f)/SO(2N_f)$
\begin{align}
  \xi(\pi)=e^{i\pi^\alpha X^\alpha/f_\pi} \ .
\end{align}
The left action of $U(2N_f)$ on $\xi$ leads to
\begin{align}
  U(2N_f):~\xi(\pi)\to g\,\xi(\pi)=\xi(\pi^\prime)h(\pi,g) \ ,~~
{}^{\exists}h\in H \ .
\end{align}
This defines the transformation law of the pion field $\pi$ to
$\pi^\prime$. The exponentiated pion field $\hU(\pi)$ is given by
\begin{align}
  \hU=\xi\xi^T \ .
\end{align}
It is easy to verify that
\begin{align}
  \hU(\pi^{\prime})=g\,\hU(\pi)g^T \ ,
\end{align}
and furthermore $U(\pi)$ is independent of the choice of the representatives
of $\xi$.

\subsection{WZW term}

The WZW term for $R$ real is obtained as follows. 
We start with the WZW term for $R$ complex with $N_f$ flavors.
Replace $N_f \to 2N_f$ yields the NG boson field
\begin{align}
  \hU\in U(2N_f) \ .
\end{align}
We impose the condition 
\begin{align}
\hU=\hU^T \ ,   
\label{UUT}
\end{align}
which is consistent with the CCWZ
construction as reviewed before.
Let $\cAb_L$ and $\cAb_R$ be the background gauge field 
for $U(2N_f)_L$ and $U(2N_f)_R$, respectively.
Then $\cAb$, that for the chiral flavor symmetry $U(2N_f)$, 
is obtained by setting
\begin{align}
  \cAb_L=\cAb \ ,~~~
  \cAb_R=-\cAb^T \ .
\label{AbLR:2Nf}
\end{align}
To see this, we note that with this
embedding, the covariant derivative for $R$ complex
(\ref{DU:Rc}) becomes that for $R$ real (\ref{DU:Rr}).
The consistency condition of (\ref{AbLR:2Nf}) with the gauge 
transformation of $\Ab_{L,R}$
\begin{align}
  \delta \Ab_{L}=d\alpha_L
+i\alpha_L\Ab_L
-i\Ab_L\alpha_L \ ,~~~
  \delta \Ab_{R}=d\alpha_R
+i\alpha_R\Ab_R
-i\Ab_R\alpha_R \ ,\nn
\end{align}
demands 
\begin{align}
  \alpha_L=\alpha \ ,~~~\alpha_R=-\alpha^T \ ,\nn
\end{align}
with $\alpha$ identified with the infinitesimal gauge transformation
parameter for the chiral $U(2N_f)$ symmetry.
Then, the WZW term for $R$ real is given by
\begin{align}
  S_{\rm WZW}^{R={\rm real}}=
\frac{1}{2}\,S_{\rm WZW}^{R={\rm cpx}}
\Big|_{N_f\to 2N_f,\,(\ref{UUT}),(\ref{AbLR:2Nf})}
\ .\nn
\end{align}
This is because 
the chiral anomaly of L-R form (\ref{dSWZW})
is evaluated as
\begin{align}
  \delta S_{\rm WZW}^{R={\rm cpx}}
\Big|_{N_f\to 2N_f,\,(\ref{UUT}),(\ref{AbLR:2Nf})}
=-\frac{2\,d(R)}{24\pi^2}\int\tr 
\left[
\alpha\, d\left(\cAb d\cAb-\frac{i}{2}\cAb^3\right)\right] \ .\nn
\end{align}
Here, tr is taken for the fundamental representation of $\mathfrak{u}(2N_f)$.

It is easy to verify that
\begin{align}
  S_{\rm WZW}^{R={\rm real}}=
-i\frac{d(R)}{48\pi^2}\int Z^{\rm r}-i\frac{d(R)}{480\pi^2}
\int\tr (\hU d \hU^{-1})^5 \ ,  
\end{align}
where
\begin{align}
    Z^{\rm r}=&~
i\,\tr\left[(\cAb d\cAb+d\cAb \cAb-i\cAb^3)
(\hU^{-1}\cAb^T\hU-i\hU d\hU^{-1})\right]\nn\\
&-\frac{1}{2}\,\tr\left[
 d\cAb\left(\hU\cAb^Td\hU^{-1}+d\hU\cAb^T \hU^{-1}\right)\right]
+i\,\tr\left[\cAb(d\hU\hU^{-1})^3\right]\nn\\
&+\frac{1}{2}\,\tr\left(\cAb\, d\hU\hU^{-1}\right)^2
+\frac{1}{2}\,\tr\left[\left(\cAb^T\hU^{-1}\cAb\hU-\hU^{-1}\cAb\hU\cAb^T\right) d\hU^{-1} d\hU\right]\nn\\
&-i\,\tr\left(\cAb d\hU\hU^{-1}\cAb \hU\cAb^T\hU^{-1}\right)
+\frac{1}{4}\tr\left(\cAb^T\hU^{-1}\cAb\hU\right)^2\ .
\label{Zr1}
\end{align}
{}Factorizing the $\eta^{\prime}$ meson as 
\begin{align}
  \hU=e^{i\wt\eta}\,U \ ,~~U\in SU(2N_f) \ ,\nn
\end{align}
and extracting from $\cAb$ the $\bZ_{4N_fT(R)}\subset U(1)_A$ gauge field
as
\begin{align}
  \cAb=\cA-\chi I_{2N_f} \ ,~~~\cA\in\mathfrak{su}(2N_f) \ ,
\nn
\end{align}
a bit lengthy computation shows that
\begin{align}
  Z^{\rm r}=
  Z^{\rm r}\big|_{\chi=\wt\eta=0}
-\int (d\wt\eta+2\chi)\,\tr \left(\cF\,UDU^{-1}\right)
+i\int d\wt\eta\,\left(3\cA d\cA-2i\cA^3\right)
+i\int \chi\,\left(4\cA d\cA-3i\cA^3\right) \ .
\label{Zr2}
\end{align}
Here, $\cF=d\cA-i\cA^2$ and 
\begin{align}
  DU^{-1}=dU^{-1}+iU^{-1}\cA+i\cA^TU^{-1} \ ,\nn
\end{align}
is the covariant derivative for the $SU(2N_f)$ gauge group.

\providecommand{\href}[2]{#2}\begingroup\raggedright\endgroup

\end{document}